# Wafer-scale growth of two-dimensional, phase-pure InSe

Seunguk Song[1], Sungho Jeon[2], Mahfujur Rahaman[1], Jason Lynch[1], Pawan Kumar[1,2], Srikrishna Chakravarthi[1], Gwangwoo Kim[1], Xingyu Du[1], Eric Blanton[3], Kim Kisslinger[4], Michael Snure[3], Nicholas R. Glavin[3], Eric A. Stach[2], Troy Olsson[1], and Deep Jariwala[1,*]

[1]Department of Electrical and Systems Engineering, University of Pennsylvania, Philadelphia, Pennsylvania 19104, United States
[2]Department of Materials Science and Engineering, University of Pennsylvania, Philadelphia 19104, United States
[3]Air Force Research Laboratory, Materials and Manufacturing Directorate, Wright-Patterson Air Force Base, Fairborn, Ohio 45433, United States
[4]Brookhaven National Laboratory, Center for Functional Nanomaterials, Upton, New York 11973, United States

[*]Correspondence should be addressed. Email to: dmj@seas.upenn.edu (D.J.)

**Abstract**

Two-dimensional (2D) indium monoselenide (InSe) has attracted significant attention as a III-VI two-dimensional semiconductor (2D) with a combination of favorable attributes from III-V semiconductors as well as van der Waals 2D transition metal dichalcogenides. Nevertheless, the large-area synthesis of phase-pure 2D InSe remains unattained due to the complexity of the binary In-Se system and the difficulties in promoting lateral growth. Here, we report the first polymorph-selective synthesis of epitaxial 2D InSe by metal-organic chemical deposition (MOCVD) over 2" sapphire wafers. We achieve thickness-controlled, layer-by-layer epitaxial growth of InSe on c-plane sapphire via dynamic pulse control of Se/In flux ratio. The layer-by-layer growth allows thickness control over wafer scale with tunable optical properties comparable to bulk crystals. Finally, the gate-tunable electrical transport suggests that MOCVD-grown InSe could be a potential channel material for back-end-of-line integration in logic transistors with field-effect mobility comparable to single-crystalline flakes.

**Introduction**

Two-dimensional (2D) semiconductors, such as indium monoselenide (InSe), have superior electrostatic control and high mobility, making them ideal for use in next-generation van der Waals (vdW)-integrated electronics[1]. InSe is particularly promising as a channel material for logic devices due to its high Hall mobility (>1,000 cm$^2$V$^{-1}$s$^{-1}$ at room temperature), a small effective mass for electrons (~0.14mo) as well as and a low asymmetry between effective masses of electrons and holes ($m_h$/$m_e$~2)[2,3]. When combined with highly p-doped Si, InSe can also exhibit steep subthreshold swing in tunnel field-effect transistors (FETs) with record high on-state current densities ($I_{on}$) and on-to-off current ratios ($I_{on}$/$I_{off}$ )[4]. Additionally, doping or applied strain in InSe layers can develop out-of-plane and in-plane ferroelectricity, offering potential for novel memory devices[5,6]. In addition, Indium being a low melting point metal[7] is expected to enable low-temperature (< 400 °C) growth of high-quality 2D semiconductors enabling vertically hetero-integrated functional layers on top of silicon logic. However, the lack of large-area growth techniques remains a challenge to realizing InSe's potential as a post-Si channel or Back-End-Of Line (BEOL) compatible channel material.

Growing high-quality 2D InSe layers has proven difficult due to the complexity of the In-Se phase diagram[8,9]. Unlike 2D group-VI TMDs or 3D III-V compound semiconductors (e.g., GaN, GaAs, InP etc.) which have only one stable polymorph, binary In-Se systems have at least four stable different structures (i.e., InSe, In$_2$Se$_3$, In$_6$Se$_7$, or In$_4$Se$_3$) at room temperature, making phase-pure synthesis challenging[9]. To achieve phase-pure and stoichiometric InSe, a delicate balance of vapor flux between In and Se is necessary. For instance, a Se-rich environment would promote the formation of In$_2$Se$_3$ rather than InSe, whereas an In-rich atmosphere may lead to a higher possibility of achieving InSe phase. On the other hand, the low chalcogen-to-metal ratio, which is necessary for InSe nucleation, can stimulate 3D-like island growth mode as opposed to 2D layer-by-layer growth[10]. Excessive In flux during the synthesis also poses a problem since it creates 3D-like In droplets, which has been observed commonly for In-based nitrides[11,12] and chalcogenides[13,14]. However, in conventional chemical vapor deposition (CVD), it is difficult to balance the flux ratio between In and Se to achieve layer-by-layer growth of 2D InSe. Additionally, using a powder-based precursor in conventional CVD decreases the source flux as a function of growth time. Vertical metal-organic CVD (MOCVD) is a potential solution for better precursor control and mass production for large-area applications[10,15]. However, epitaxy of InSe on an insulating substrate has not been explored, despite the potential for well-oriented polydomains to improve structural and electronic properties.

In this work, we present a successful approach for growing a polymorph-selective, high-quality, and thickness-controlled 2D InSe thin film through vertical, cold-walled metal-organic chemical vapor deposition (MOCVD). By periodically interrupting the Se source during MOCVD, we create a Se-deficient environment that favors the nucleation of InSe over $In_2Se_3$. This pulsed delivery of the Se precursor also suppresses In-rich droplet formation and facilitates the lateral growth of InSe at low growth temperatures (360-500 °C), allowing us to achieve a highly stoichiometric, crystalline thin film on c-plane oriented 2-inch sapphire. These growth temperatures are also suitable for back end of line (BEOL) integration in leading edge Si microelectronics, further enhancing the application appeal of our results. The resulting single-crystalline InSe domains are directionally oriented along the crystal structure of the substrate, and the thickness of InSe can be controlled as a function of growth time. We also demonstrate few-layer InSe transistors with a high on-to-off current ratio ($I_{on}/I_{off}$ ratio $\approx 10^4$-$10^5$) and two-terminal field-effect mobility ($\mu_{FE} \approx 2.8$ $cm^2V^{-1}s^{-1}$), comparable to those of mechanically exfoliated single crystals. Our work represents a promising step towards constructing phase-pure 2D InSe films on a wafer scale using a method that can be adapted to other material systems with various polymorphs.

**Results and discussion**

To synthesize 2D InSe layers, we perform MOCVD in a cold-walled, vertical reactor with gas-phase precursors of trimethylindium [TMIn; $(CH_3)_3In$] and dimethyl selenide [DMSe; $(CH_3)_2Se$] as the In and Se sources, respectively (**Figure 1**a). The individual pressure controllers and mass flow controllers allow subtle control of molecular flow rates of each precursor, which dominantly impacts the resulting polymorphic crystal in the binary In-Se system (Figure 1b and Figure S1). For example, for a molecular flow rate of Se higher than In (i.e., high Se/In ratio), the growth typically results in $In_2Se_3$ crystal (Figure 1b, top) since $In_2Se_3$ has one more Se atomic layer in between two In-Se layers of InSe crystal (Figure 1b, bottom). The vertical delivery of precursors from the showerhead to the rotating substrate in the system (Figure 1a) also supported the homogenous nucleation and growth of the 2D crystals across the 2-inch wafer (Figure 1c), which is a significant benefit over horizontal CVD[16]. See methods for additional details on precursors, reactor, and growth. The *c*-plane sapphire serves as a substrate for van der Waals epitaxy because of its hexagonal lattice symmetry matching with InSe, as explained later.

More importantly, to expand the InSe growth window, we varied the flow rate of DMSe as a function of growth time (*t*), as illustrated in Figure 1d. We observe that a pulse-like DMSe flow promotes InSe nucleation rather than $In_2Se_3$ due to the momentarily induced Se deficiency in the growth environment. In Figure 1e, we plot the growth phase diagram showing the resultant InSe and $In_2Se_3$ synthesized depends strongly on the maximum Se/In ratio and the fraction of the unit delivery time of DMSe ($t_d$) in the total pulse time ($t_i+t_d$) (i.e., Max Se/In ratio *vs.* the value of $t_d/(t_i+t_d)$). When the DMSe flow rate is held constant ($t_d/(t_i+t_d) = 1$, as $t_d = 0$), the growth window for InSe is narrow; for instance, only for Se/In ratio below ~2 results in the production of the InSe islands (Figure S2) owing to the Se-rich condition. Furthermore, the InSe growth under constant DMSe flow with a smaller Se/In ratio of ~2 shows a droplet-like (not 2D) morphology due to excess In flux during the MOCVD (Figure S2b) which leads to formation of metallic In or In-rich InSe islands. On the other hand, the pulsed delivery of DMSe (i.e., $t_d/(t_i+t_d) < 1$) widens the growth window for the InSe due to the periodically induced In-rich growth environment which is favorable for the InSe nucleation. Further, we observe that even for a high Se/In ratio of ~11 in a pulse mode ($t_d/(t_i+t_d) < 1$) resulted in the synthesis of atomically thin InSe layer (Figure S3). In addition, the pulse-like injection of precursor promotes the ripening of the nuclei and their lateral growth to coalesce into continuous high-quality 2D InSe, similar to the previous reports for TMDs[17,18]. To substantiate this further, we have characterized the crystalline quality and growth morphology (e.g., 2D or 3D growth modes and formation of In-rich droplets) as a function of $t_d$ and $t_i$ values (See Supplementary Note 1 and Figure S4).

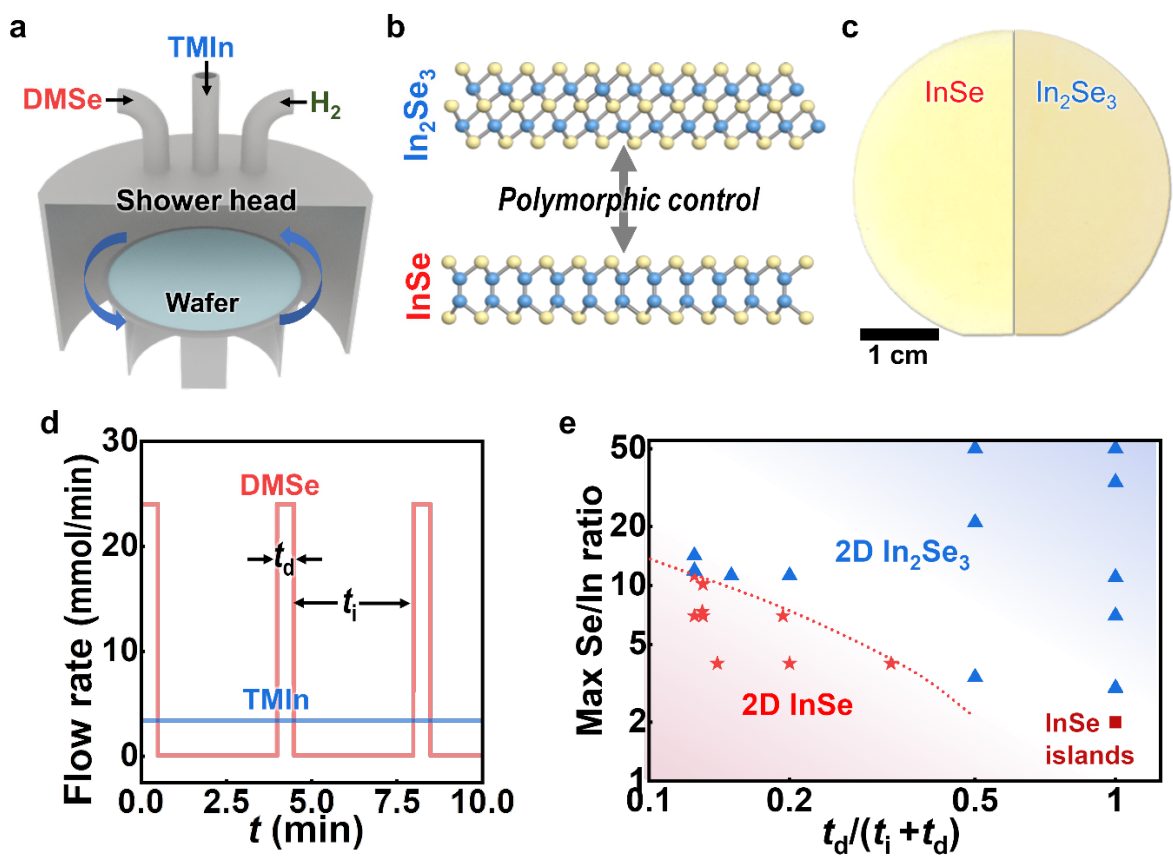

**Figure 1. Large-scale metal-organic chemical vapor deposition (MOCVD) of indium selenides polymorphs.** (**a**) Schematic of vertical MOCVD used in this study. The dimethyl selenide [DMSe; $(CH_3)_2Se$] and trimethylindium [TMIn; $(CH_3)_3In$] are used as Se and In precursors, respectively, which are delivered by the bubbling system. (**b**) Atomic structures of indium selenide polymorphs, e.g., $\beta$-phase $In_2Se_3$ (top) and InSe (bottom). (**c**) Photographs of the monolayer InSe (left) and $In_2Se_3$ (right) deposited on 2-inch $c$-plane sapphire. (**d**) Molecular flow rate of DMSe (red) and TMIn (blue) precursors as a function of growth time ($t$) at the growth temperature of 500 °C. The delivery and interruption time of DMSe during one pulse period is labeled as $t_d$ and $t_i$, respectively. (**e**) Phase diagram of $\beta$-$In_2Se_3$ and InSe structures depending on the Se/In ratio and the $t_d/(t_i+t_d)$ value during the growth at the temperature of 500 °C. The $t_d/(t_i+t_d)$ value is 1 when the DMSe constantly flows ($t_d = 0$).

We identify the resulting 2D polymorphs by using Raman spectroscopy, since the representative signals from InSe (red) and In$_2$Se$_3$ (blue) are distinct as seen in **Figure 2**a. Different peaks are displayed depending on the crystals, with three $A_1$ vibrational modes for $β$-phase In$_2$Se$_3$ and two $A_{1g}$ and $E_{2g}$ modes for InSe. A slightly large full-width at half maximum (FWHM) of ~11 cm$^{-1}$ of the first $A_1$ peak at ~108.5 cm$^{-1}$ of $β$-In$_2$Se$_3$ is caused by the two mixed longitudinal (LO) and transverse (TO) components, not poor crystalline quality[19]. Raman mapping of the InSe film grown by pulsed DMSe delivery is also performed (Figure S3c-f) which shows the uniformity over the large area of the film. Characteristic InSe peak positions, uniform intensities, and FWHM are visible across a large area (~100 μm) without any traces of In$_2$Se$_3$ (Figure S3c, f). The large-scale homogeneity of the monolayer InSe film grown on 2-inch sapphire is also indicated by the uniform contrast in both the photograph (Figure 1c) and the corresponding optical microscopy image (Figure S3).

Analysis based on X-ray photoelectron spectroscopy (XPS) for In $3d$ and Se $3d$ scan also suggests the synthesis of pure phase InSe rather than In$_2$Se$_3$ (Figure 2b, c). The In $3d_{3/2}$ and In $3d_{5/2}$ core levels of InSe are located at lower binding energies compared to those of In$_2$Se$_3$ due to the higher positive oxidation state of the compounds (Figure 2b). On the other hand, the Se $3d_{3/2}$ and $3d_{5/2}$ of In$_2$Se$_3$ are located at lower energy levels than those of InSe because of the loss of valance charge of the Se atom[20] (Figure 2c). The absence of Se-O bonds at ~58.6 eV[21] implies that MOCVD produces high-quality InSe or In$_2$Se$_3$ without oxide impurities. The XPS-driven atomic ratio between Se and In (i.e., at. % of Se/In) of the as-grown film is ~1.03, which is similar to the ideal stoichiometry of InSe (i.e., 1.00). In contrast, the at. % of Se/In of as-grown In$_2$Se$_3$ thin film is higher (~1.30).

The growth of phase-pure InSe by the precursor-interrupted MOCVD is found to be universal as the selection of InSe and In$_2$Se$_3$ nucleation relies highly on the flow rate control (as plotted in Figure 1e). Specifically, InSe is synthesized with the optimized recipe (see Methods) regardless of the target substrates (i.e., amorphous SiO$_2$/Si, AlO$_x$/Si, and sapphire) as suggested by Raman spectra (Figure S5a-c) and X-ray diffraction (XRD) patterns (Figure S5d-f). The X-ray diffraction patterns display (002$l$) peaks of the crystals matching with the JCPDS database of InSe (card

No.:34-1431, where *a* and *c* is 4.005 Å and 16.640 Å, respectively), which is indicative of *c*-plane oriented InSe on arbitrary substrates (Figure S5). The synthesis of InSe with (002*l*) texture is also possible at lower temperatures of ~360 or 400 °C (Figure S6), indicating the scalability of the method to directly integrate with various electronic components at temperatures compatible with complementary metal-oxide semiconductor (CMOS) back-end-of-line processes.

The thickness control of 2D InSe thin film on a large scale is facilitated by adjusting the growth time, attributed to its layer-by-layer growth mode (Figure 2d-g). In particular, the small total flow (< 22 sccm) towards a low growth rate enables InSe to exhibit its growth aspects as a function of growth time (*t*). For thin films grown for a growth time *t* of ~35 min ($\approx 0.8t_0$), small domains of monolayer InSe are observed in the AFM image (Figure 2d). A fully stitched monolayer InSe film is found when the growth time is ~45 min (= $1.0t_0$, and $t_0$ denotes the time *t* when a continuous monolayer film formed; Figure 2e). A growth time of $1.2t_0$ allows for the further nucleation of InSe domains on the pre-formed monolayer InSe film (Figure 2f). A much longer growth time of ~$1.9t_0$ results in the synthesis of additional multilayers of InSe, where the edges of the single domains tend to align along one direction (Figure 2g and Figure S7a, b). This signifies that the newly formed crystalline layers are synthesized coherently with well-defined orientations with respect to the underlying substrate (i.e., *c*-plane sapphire), which serves as important evidence of epitaxial growth. The thickness of the thin films grown for $1.9t_0$ and $2.7t_0$ are estimated to be ~5-6 and ~11 nm (i.e., ~6-7 and ~13-14 layers), respectively, with an estimated vertical growth rate of ~$0.133 \pm 0.002$ nm/min (~8 nm/h obtained from the slope of linear fit in Figure 2h), which is considerably faster than the growth rates of other MOCVD of group-VI 2D sulfides[22,23].

Optical characterization of crystals with different thicknesses is performed using 2D InSe synthesized at different growth times, as represented by the representative Raman and photoluminescence (PL) spectra in Figure 2f, g. The Raman spectra of the InSe grown for $2.7t_0$ (with a thickness of ~11 nm) show the Raman vibrational modes of $A^1_{1g}$, $E_{2g}$, and $A^2_{1g}$ at ~112.9, 176.7, and 224.9 cm$^{-1}$, respectively. As growth time and corresponding thickness decrease, the position difference between out-of-plane modes ($A^1_{1g}$ and $A^2_{1g}$) becomes bigger (Figure 2i and Figure S8a, b) due to variations in interlayer force, which is consistent with literature[24]. Despite the drastic decrease in Raman intensities in the monolayer (grown for $1.0t_0$) as expected[24,25], the uniformity of the film is indicated by Raman mapping (Figure S3c). The PL spectra of the InSe film prepared at different growth times, normalized to the maximum peak intensities, also demonstrate thickness dependence (Figure 2j). Since the InSe films below ~7 layers have an indirect band gap[26], the intensity of PL emission from film grown for 1.0-1.9$t_0$ is far smaller than that of 2.7$t_0$ (~11 nm) (Figure S8c). Furthermore, the excitonic transition energy of the layers is distinct depending on the growth time due to thickness-dependent quantum confinement[25,26] (Figure 2g and Figure S8d). The PL emission of the as-grown InSe film indicates the preparation of crystals without substantial disorders or gap states. In the case of the as-grown InSe monolayer, no distinct PL signal is detected due to its indirect transition, air-induced oxidation, or high photon energy (~2.0 eV) to be observed using a 633 laser. However, band-to-band transitions in the as-grown InSe monolayer are apparent in the reflectance spectra and the ellipsometer measurements for the complex refractive index ($\eta = n + ik$), indicating its characteristic band structure of the high-quality 2D structure (Figure S9).

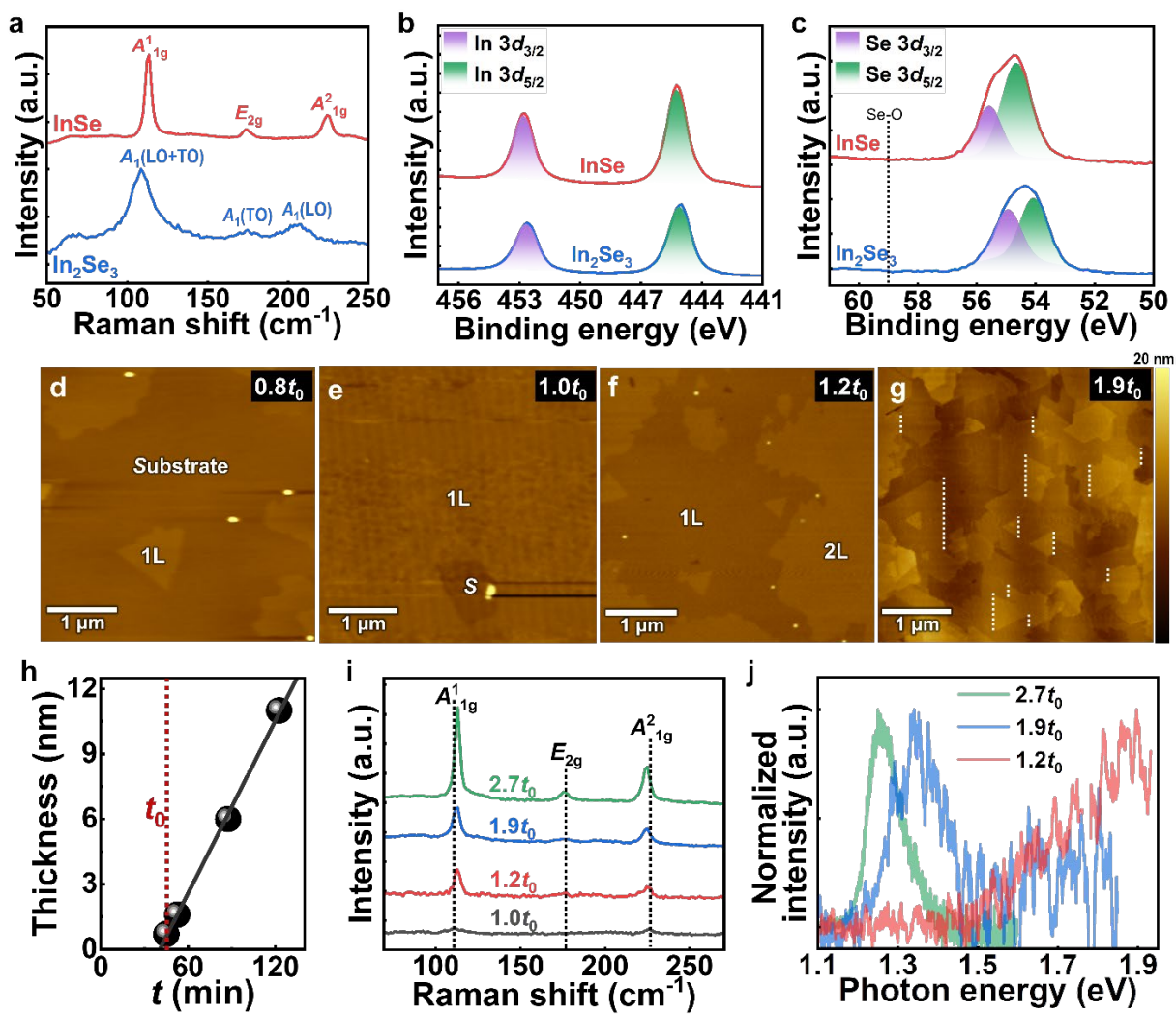

**Figure 2. Polymorph-selective and layer-by-layer synthesis of InSe layers.** (**a**) Representative Raman spectra of InSe (red) and $In_2Se_3$ (blue) prepared in this study. (**b, c**) The X-ray photoelectron spectroscopy (XPS) analysis of InSe (red) and $In_2Se_3$ (blue) with respect to the (**a**) In $3d$ scan and (**b**) Se $3d$ scan. (**d-g**) AFM images of the InSe thin film grown for different growth times, $t$; (**d**) $0.8t_0$, (**e**) $1.0t_0$, (**f**) $1.2t_0$, and (**g**) $1.9t_0$, where the $t_0$ denotes the time for growing fully stitched monolayer InSe on sapphire. The dashed line in (g) highlights one edge of the InSe domains, indicating the arranged crystal orientations. (**h**) Summary of the thickness ($H$) of the InSe as a function of the growth time ($t$). The slope of the linear fit specifies the vertical growth rate, ~0.133 ± 0.002 nm/min. (**i**) Raman spectra of the as-grown InSe thin films depending on the growth $t$; e.g., $1.0t_0$, $1.2t_0$, $1.9 t_0$, and $2.7t_0$. The dashed lines indicated the $A^1_{1g}$, $E_{2g}$, and $A^2_{1g}$ modes of the InSe monolayer grown for $1.0t_0$. (**j**) Photoluminescence (PL) spectra of the InSe film synthesized via the MOCVD with different $t$. The spectra were normalized to their maximum peak after subtracting the background noise from the sapphire substrate.

Scanning transmission electron microscopy (STEM) is conducted to observe the crystal structure of few-layer InSe synthesized by repeated interruptions of DMSe flow (**Figure 3**). The structural investigation of an InSe cross-sectioned along the zigzag direction ($[11\bar{2}0]_{InSe}$) is first performed in Figure 3a-c. The high-angle annular dark field (HAADF) STEM image in Figure 3a shows the layer-by-layer structure of InSe with clearly visible vdW gaps. The interlayer distance along the [0001] direction is measured as ~0.83 nm, which is consistent with the AFM measurement of the monolayer (Figure 2d-g and Figure S3b) and the previous report for InSe[3]. The atomic configuration of Se-In-In-Se resolved in STEM (inset of Figure 3a and Figure 3c) indicates the ~5-6 layers of InSe instead of $In_2Se_3$. The corresponding selected-area diffraction pattern (SAED) in Figure 3b suggests that the single-crystalline nature of InSe grown on sapphire (0001) while implying the epitaxial relation of $(0001)_{InSe}//(0001)_{Al_2O_3}$ and $(10\bar{1}0)_{InSe}//(11\bar{2}0)_{Al_2O_3}$. The interlayer spacing of $(10\bar{1}0)_{InSe}$ and $(11\bar{2}0)_{InSe}$ extracted from the atomic-resolution STEM images in Figure 3c, d (and Figure S10a, c) is ~0.35 nm and ~0.20 nm, respectively, which is in good agreement with those in previously reported 2D InSe structures[27,28]. The plain-view STEM images (Figure 3e and Figure S11) and the corresponding Fourier transformed pattern (inset of Figure S11a) also imply the high crystallinity of 2D InSe with three-fold symmetry along the [0001] zone axis. The high quality of InSe is additionally evidenced by the TEM-energy dispersive X-ray spectroscopy (EDS) (Figure S12) and Raman spectra (Figure S13), which reveal the almost perfect stoichiometric property of InSe without carbon-related impurities.

In conjunction with the SAED (Figure 3b), the extracted atomic distances of $Al_2O_3$ planes and the coordinated InSe structures in the atomically resolved STEM images (Figure 3c-e) suggest that the InSe zigzag direction is perpendicular to the $[10\bar{1}0]$ zone axis of sapphire (Figure 3c) and the armchair direction is parallel to the $<10\bar{1}0>_{Al_2O_3}$ (Figure 3d) (see Figure S10 for more STEM images and details on the relation of $(11\bar{2}0)_{InSe}//(10\bar{1}0)_{Al_2O_3}$). The orientation-controlled InSe on the *c*-plane sapphire implies that epitaxy is possible by following the correlation presented in cross-sectional and top-view atomic illustrations of Figure 3f, g. Accordingly, the epitaxially grown InSe grains can merge together during MOCVD to form small-angle grain boundaries (GBs) or nearly single crystals, as directionally aligned InSe domains are observed in the AFM images (Figure 2g and Figure S7). Further, no atomic discontinuity such as GB is observed during cross-sectional

STEM analysis of the InSe along the zigzag direction over several μm length of the scam/imaging. The epitaxial growth mode of InSe also evokes the possibility of obtaining unidirectional InSe single crystals using various sapphire wafers cut along different crystallographic orientations in the future.

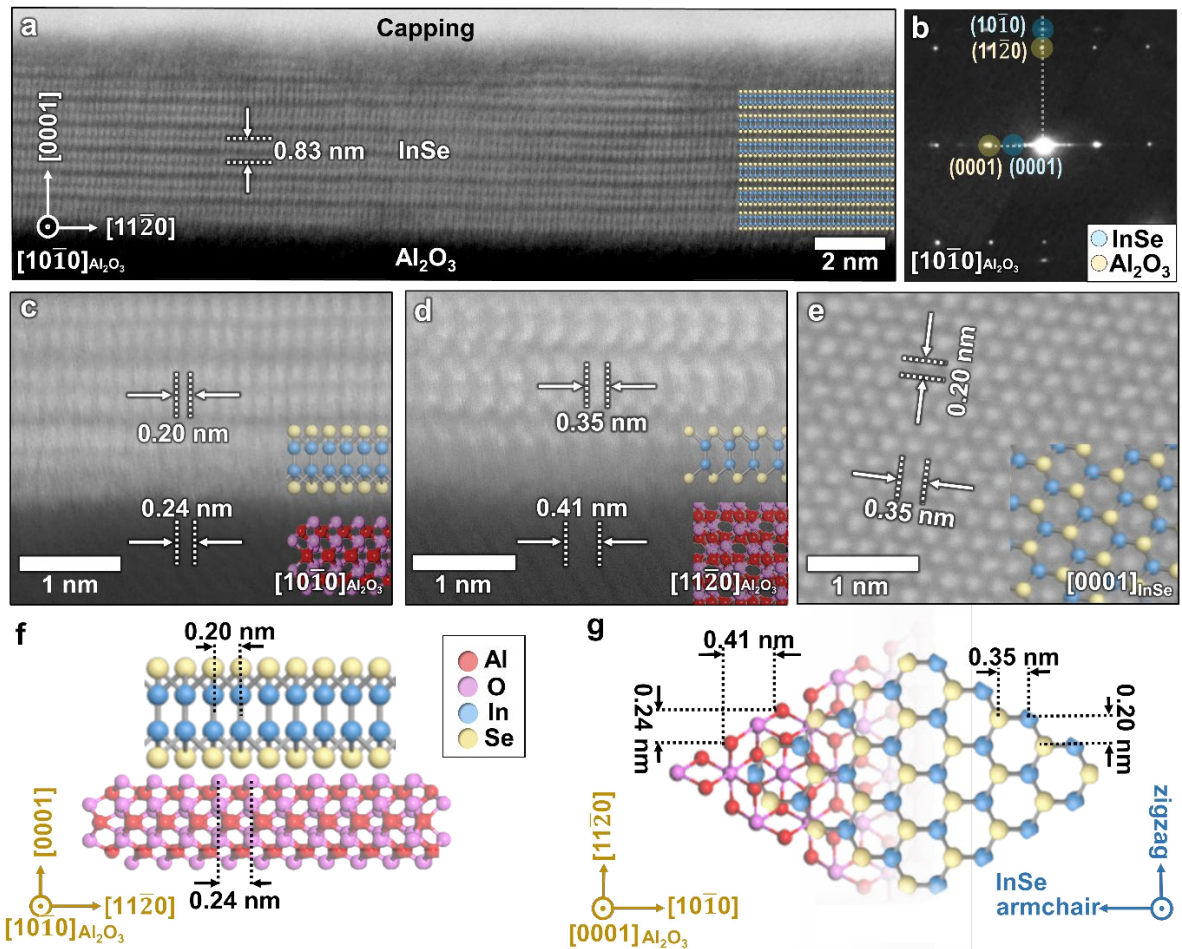

**Figure 3. Epitaxial relationship between synthetic InSe and sapphire (0001) substrate.** (**a**) Cross-sectional scanning transmission electron microscopy (STEM) image of the few-layer InSe. The capping layer of a metal, Ir, was deposited to minimize unexpected damage during its cross-sectioning process. (**b**) Corresponding selected-area diffraction patterns of heterointerface in (a), displaying the epitaxial relation between InSe on $Al_2O_3$ along the $[10\bar{1}0]_{Al_2O_3}$ zone axis. (**c-e**) Atomic resolution STEM images of the InSe captured along the zone axis of (**c**) $[10\bar{1}0]_{Al_2O_3}$, (**d**) $[11\bar{2}0]_{Al_2O_3}$, and (**e**) $[0001]_{InSe}$. The atomic illustrations in (c-e) depict each InSe and $Al_2O_3$ plane. The schematic in (e) reveals the atomic structure of monolayer InSe. (**f**) Side and (**g**) top views of atomic schematics illustrating the epitaxial relation of InSe on a sapphire (0001) surface.

The electrical transport of the grown InSe films is characterized by two-terminal FET devices at room temperature (**Figure 4**). Device fabrication is conducted using a top-gate geometry with high-$k$ dielectric layers (i.e., $Sb_2O_3/HfO_2$)[29] to avoid a wet-transfer process that can degrade the channel by severe oxidation[3] (Figure 4a and Figure S14a). The FETs manufactured based on few-layer InSe (grown for 1.9$t_0$) demonstrate gate-modulated electrical conductance, as indicated in their representative transfer and output curves (Figure 4b, c and Figure S14b). A $I_{on}/I_{off}$ value of ~$10^4$-$10^5$ is achieved in FETs with a gate length of ~2 μm, which is the highest value among the reported synthesized InSe-based FETs[30-33]. Furthermore, the best two-terminal $\mu_{FE}$ of our few-layer InSe device approaches ~2.8 $cm^2V^{-1}s^{-1}$ (and ~1.0 ± 0.8 $cm^2V^{-1}s^{-1}$ averaged for more than 10 devices in Figure S14c), higher than most literature values (~0.2-3.0 $cm^2V^{-1}s^{-1}$) for exfoliated[34,35] or directly grown InSe[31,32] with thicknesses below 10 nm, as summarized in Figure 4d (and Table S2). Notably, it is also found that our monolayer InSe FETs exhibit $\mu_{FE}$ (~0.01 $cm^2V^{-1}s^{-1}$) and $I_{on}/I_{off}$ ratio (~$10^2$) (Figure S15), in which the properties are similar to that of mechanically exfoliated single-crystalline monolayer[3]. The high $\mu_{FE}$ of our few-layer and monolayer InSe implies that the GB scattering of carriers is negligible, probably contributed by the well-oriented polycrystalline domains with small-angle GBs of the epitaxially-grown InSe. The significant electrical property degradation in the monolayer compared to the few-layer InSe is ascribed to more air instability and severe carrier scattering at the substrate interface in monolayer InSe[3,34,35].

We further note that the MOCVD-grown, atomically-thin InSe is found to substantially suffer from air degradation in comparison to bulk InSe, indicating that further improvement of device performance could be possible through encapsulation using passivation layers[3]. Additionally, optimization of contact electrodes on InSe would be beneficial in realizing even higher $\mu_{FE}$, as the operation is significantly affected by Schottky barrier height and contact resistance at the metal-semiconductor junction[36]. For instance, in our experiments, the In/Au contact electrode resulted in better transport behavior than those with Ti/Au or Au (Figure S14c-f), suggesting that further investigation of the contact interface would allow for efficient carrier injection. Therefore, in combination with the integration scalability on various substrates (Figure S6), the electronic quality of the MOCVD-grown vdW InSe suggests that our synthesis process is promising for the fabrication of high-performance electronic devices based on the III-VI compound semiconductor for future heterointegration in high-performance microelectronics.

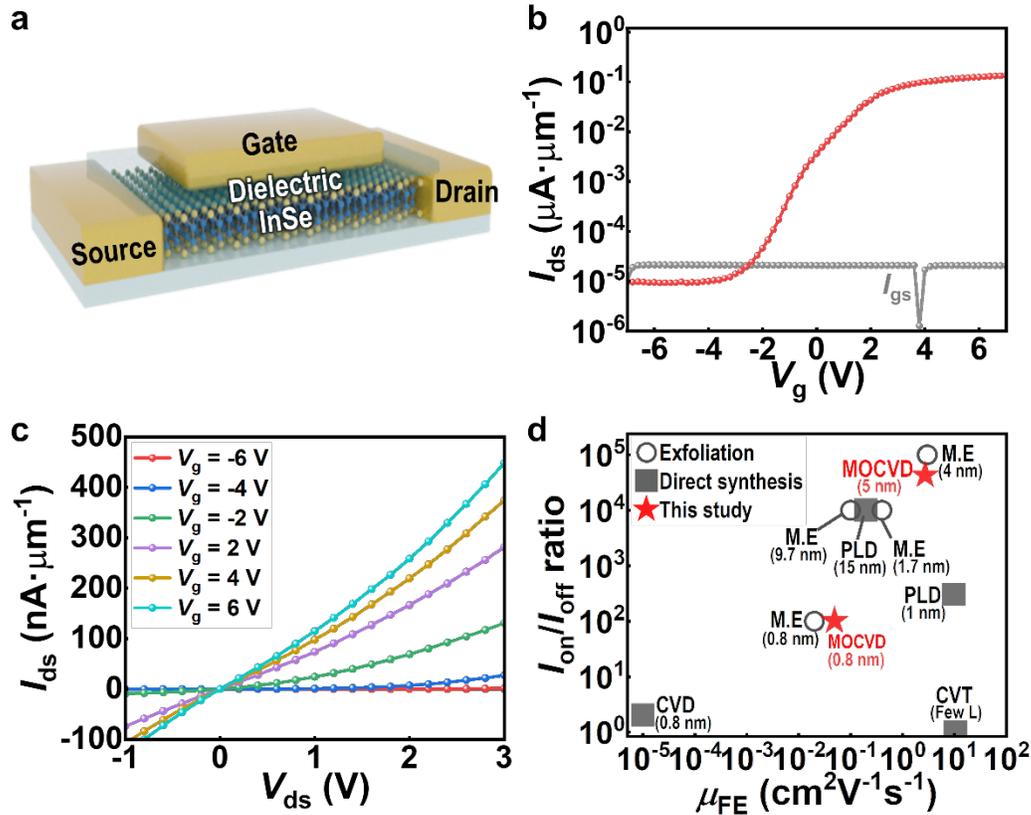

**Figure 4. Room-temperature electrical properties of few-layer InSe field-effect transistors (FETs).** (**a**) Schematic depicting the fabricated 2D InSe-based FET with a top-gate geometry. (**b**) Drain current-gate voltage ($I_{ds}$-$V_g$) curves of FETs based on few-layer (~5-6 nm) InSe, when the drain-source voltage ($V_{ds}$) of 1 V was applied. (**c**) Output ($I_{ds}$-$V_{ds}$) characteristics for FETs based on few-layer InSe film under the different $V_g$. (**d**) Comparisons of the field-effect mobility ($\mu_{FE}$) and on-to-off current ($I_{on}/I_{off}$) ratio of our two-terminal InSe FET with previously reported values of devices prepared by mechanical exfoliation (M.E)[3,34] and direct synthesis (i.e., pulsed-laser deposition followed by annealing (PLD)[30,31], chemical vapor transport (CVT)[32] and chemical vapor deposition (CVD)[33].). For fair benchmarking, the FETs fabricated with a solid-phase dielectric layer are compared. The thicknesses of the channel in references[3,30-34] are noted in each parenthesis, which is mostly below 10 nm. See Table S1 for more details on the comparisons.

**Conclusions**

In summary, we demonstrate the growth of thickness-controlled, phase-pure InSe atomic layers via MOCVD using pulse-like injection of Se precursor DMSe. The pulse-like injection of DMSe is found to create suitable growth conditions for 2D InSe nucleation and its lateral growth. The epitaxial growth of high-quality, highly stoichiometric InSe thin film on *c*-plane sapphire is confirmed through structural and optical characterizations. Furthermore, the two-terminal FETs fabricated using few-layer InSe are found to validate the high $I_{on}/I_{off}$ of ~$10^4$-$10^5$ and $\mu_{FE}$ of ~2.8 cm$^2$V$^{-1}$s$^{-1}$ which are comparable to those of mechanically exfoliated single-crystalline flakes with a similar thickness. In addition, the optical properties of the grown InSe are found to be comparable to CVT-grown bulk crystals. Our work is the first demonstration of phase pure epitaxial growth of InSe and offers novel insights into the vapor phase growth of polymorph-complex binary 2D chalcogenide systems. Our work is expected to open up new directions in production of wafer-scale electronic-grade 2D III-VI compound semiconductors at low temperatures (360-500 °C).

**Methods**

**MOCVD growth of InSe and In$_2$Se$_3$.** InSe was grown in a showerhead-type MOCVD reactor depicted in Figure 1a. Gas-phase TMIn and DMSe were used as In and Se precursors, respectively, and delivered into the chamber by using H$_2$ as a carrier gas. The growth was mostly conducted at the *T* of ~500 °C because the pyrolysis of the TMIn and DMSe sources is effective above 480 °C[37] and 450-500 °C[38], respectively. The total pressure of 100 torr was used for the optimized growth run. For the repeated interruption of the DMSe precursor, the $t_d/(t_i+t_d)$ value of ~0.14 with a maximum Se/In ratio of ~10 (depicted in Figure 1d) was designated. The H$_2$ flow of 10 sccm was constantly injected into the chamber during the growth. Depending on the total *t*, the thickness of InSe was varied, but the fully stitched monolayer thin was obtained for a total *t* of ~45 min. In the case of In$_2$Se$_3$ growth, constant flow of DMSe with the Se/In ratio of 10 was used. The 2-inch Al$_2$O$_3$(0001) (UniversityWafer) was mainly used as the growth substrate throughout the study due to its surface smoothness and hexagonal crystal structure that matched that of the in-plane InSe structure; nonetheless, the synthesis of 2D InSe on the arbitrary substrate was also possible (Figure S7).

**Structural characterizations.** Raman spectra, PL, and reflectance spectra were measured under air ambient using a Horiba LabRam HR Evolution Confocal Microscope with a 100× objective lens (Olympus) and a 633 nm laser with a power of ~5.5 × $10^8$ mW·cm$^{-2}$ and ~0.5 μm spot size. The resonant $A_2$"-LO Raman peak at ~200 cm$^{-1}$ was absent in our InSe measurement because the used excitation laser of 633 nm resulted in the non-resonant Raman process[25]. The morphology of InSe was characterized by AFM (OmegaScope-R (AIST-NT)). STEM images and STEM-EDS were acquired using aberration-corrected STEM (JEOL NEOARM). We used a 4 cm camera length and a 6C probe size were used with a 40 μm condenser aperture, yielding approximately 62.5 pA probe current and a convergence angle of approximately 27 mrad. TEM images and SAED patterns were acquired by JEOL JEM-F200 with OneView camera. Both TEM instruments were operated at 200 kV. The cross-sectional sampling was conducted using Xe$^+$ plasma-based focused-ion-beam (FIB) system (TESCAN S8000X FIB/SEM). XPS was characterized using an SSI Spectrometer under a high vacuum (~10$^{-9}$ torr) with monochromated Al $K_a$ radiation (1486.6 eV), in which the data were corrected using the C signals.

**Electrical device fabrications and measurements.** The electrical devices were fabricated using electron-beam lithography (Elionix ELS-7500EX). Polymethyl methacrylate (PMMA) was used as an e-beam resist. The In/Au (10/40 nm) metal contacts were deposited by a thermal evaporator (Kurt J. Lesker Nano-36). The gate dielectric layers of Sb$_2$O$_3$ (10 nm) and HfO$_x$ (10 nm) were deposited by thermal evaporator atomic-layer deposition, respectively[29], followed by the deposition of gate electrodes (Ti/Au (10/40 nm)) by e-beam evaporator (Kurt J. Lesker PVD-75). The electrical measurements were conducted at room temperature by a probe station (Lake Shore) combined with a semiconductor analyzer (Keithley 4200).

**Author information**

**Contributions.** S.S. performed most of the experiments with assistance from S.J., M.R., J.L., S.C., and G.K.; S.J., P.K., K. K. and E. S. performed the high-resolution TEM imaging of the samples;

X. D. and T. O. performed the XRD characterizations; E. B. and K. K. performed the XPS measurements; S.S. and D.J. wrote the manuscript; all the authors revised and commented on the manuscript; D.J. conceived, planned and supervised the project. All authors contributed to the writing of manuscript and interpretation of the data.

**Corresponding authors.** Correspondence to Deep Jariwala.

**Acknowledgements.** D.J. and R.H.O acknowledge primary support from DARPA TWEED program HR0011-21-9-0055. S.S. was primarily supported by Basic Science Research Program through the National Research Foundation of Korea (NRF) funded by the Ministry of Education (Grant No. 2021R1A6A3A14038492). (S)TEM data and images were collected in the Singh Center for Nanotechnology at the University of Pennsylvania, supported by the Center for Hybrid Approaches in Solar Energy to Liquid Fuels (CHASE), an Energy Innovation Hub funded by the U.S. Department of Energy, Office of Science, Office of Basic Energy Sciences under Award Number DE-SC0021173.

**Supplementary information**

Figures S1-15, Table S1 and Note S1.

# Supplementary information for:

## Wafer-scale growth of two-dimensional, phase-pure InSe

Seunguk Song[1], Sungho Jeon[2], Mahfujur Rahaman[1], Jason Lynch[1], Pawan Kumar[1,2], Srikrishna Chakravarthi[1], Gwangwoo Kim[1], Xingyu Du[1], Eric Blanton[3], Kim Kisslinger[4], Michael Snure[3], Nicholas R. Glavin[3], Eric A. Stach[2], Troy Olsson[1], and Deep Jariwala,[1]

[1]*Department of Electrical and Systems Engineering, University of Pennsylvania, Philadelphia, Pennsylvania 19104, United States*
[2]*Department of Materials Science and Engineering, University of Pennsylvania, Philadelphia 19104, United States*
[3]*Air Force Research Laboratory, Materials and Manufacturing Directorate, Wright-Patterson Air Force Base, Fairborn, Ohio 45433, United States*
[4]*Brookhaven National Laboratory, Center for Functional Nanomaterials, Upton, New York 11973, United States*

[*]Correspondence should be addressed. Email to: dmj@seas.upenn.edu (D.J.)

- Figures S1-S15
- Tables S1
- References 1-22

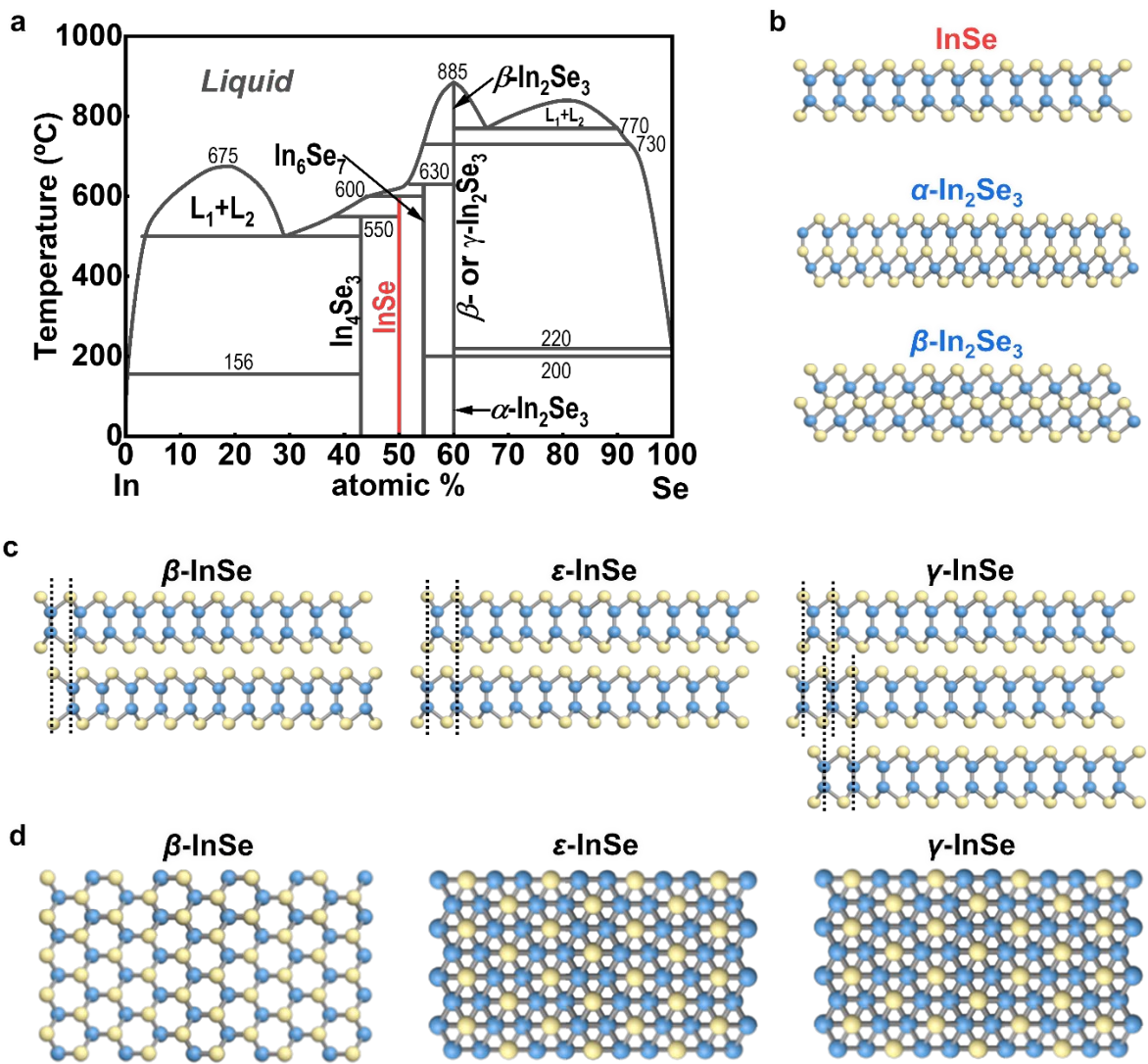

**Figure S1. Polymorphic crystalline structures of InSe and In₂Se₃.** (**a**) In-Se binary phase diagram[1]. (**b**) Atomic schematics of 2D InSe, and In2Se3 monolayers with *α* and *β* phase. (**c**) cross-sectional and (**d**) top-view atomic schematics depicting the stacking order difference in few-layer InSe with different phases.

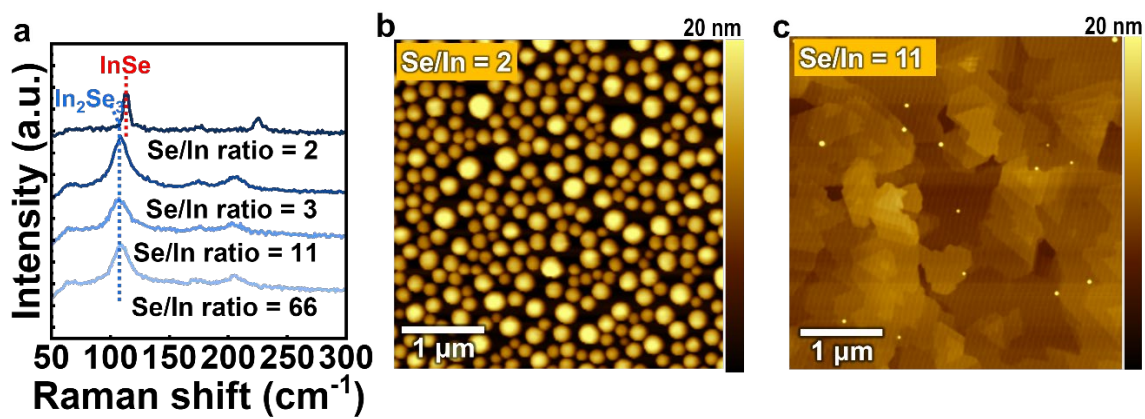

**Figure S2. Characterizations of the InSe and In₂Se₃ crystals grown under the constant flow of the DMSe.** (**b**) Evolution of Raman spectra of InSe depending on the Se/In ratios (~2-66) under the constant flow of the DMSe ($t_i$ = 1 in Figure 1c). (**c, d**) AFM images of the nanostructures prepared by the Se/In ratio of (**b**) 2 and (**c**) 11. The Se/In ratio above 2 resulted in the growth of In₂Se₃ instead of InSe. See Supplementary Note 1 for more details.

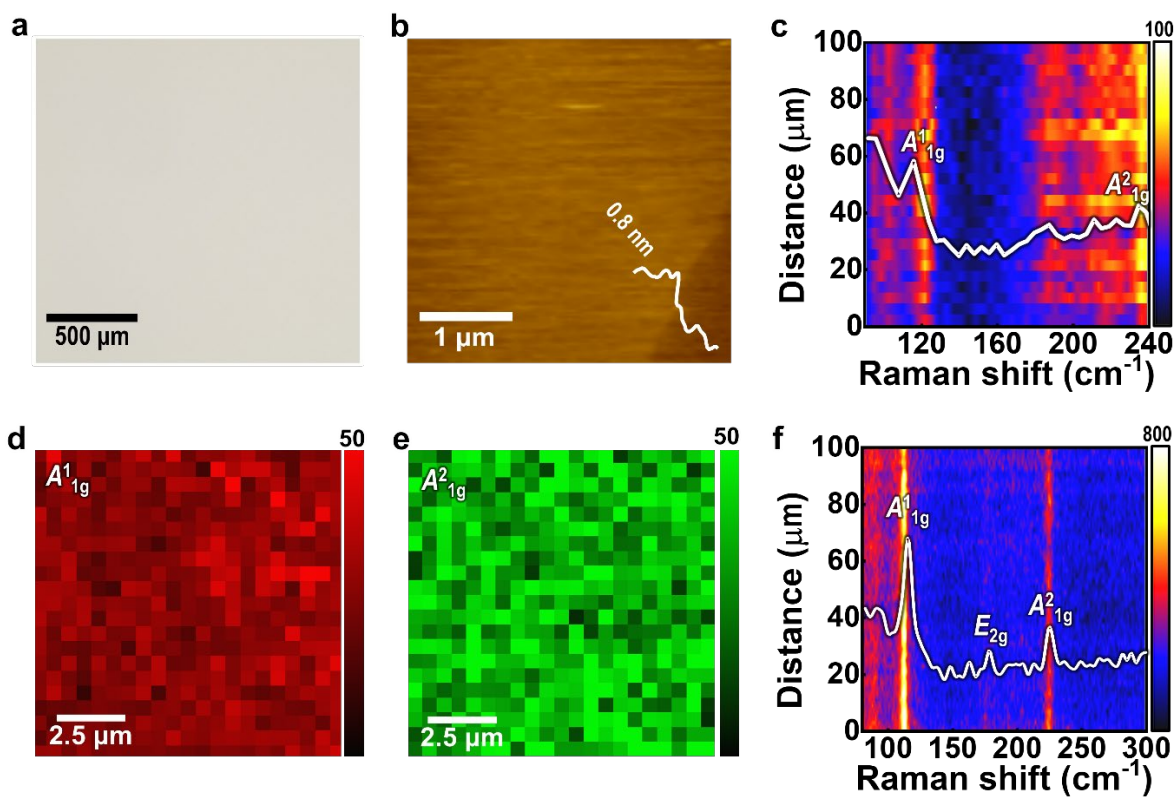

**Figure S3. Homogeneity of monolayer and bilayer InSe thin film.** (**a**) Representative optical microscopy (OM) image of the monolayer thin film. (**b**) Atomic force microscopy (AFM) image of 2D InSe, indicating the monolayer with a thickness of ~0.8 nm. (**c-e**) Representative Raman characteristics of the monolayer InSe thin film grown for $1.0t_0$. (**c**) Color-coded rendering of the spatially resolved Raman spectra. The data were collected from 100 positions with an interval of 2 μm. (**d, e**) The Raman mapping images of the InSe monolayer conducted for (**c**) $A^1_{1g}$ and (**d**) $A^2_{1g}$ peaks. Note that the achievement of the larger-area InSe monolayer mapping was challenging due to the real-time oxidation of InSe during measurement under air. (**f**) Raman mapping image for a bilayer InSe film. The data were collected from 100 positions with an interval of 1 μm. The inset shows a representative Raman spectrum of the corresponding InSe structure.

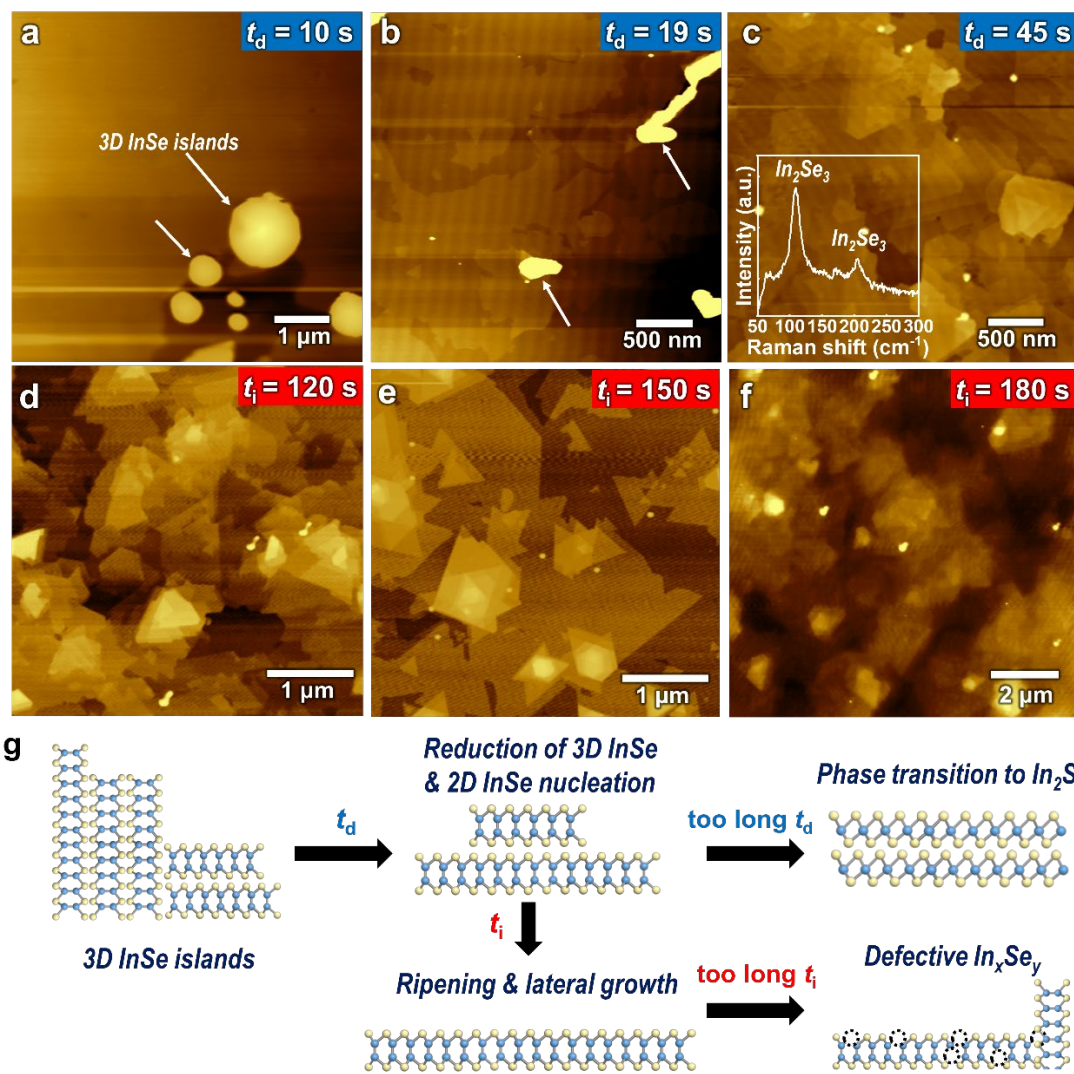

**Figure S4. Effects of DMSe delivery ($t_d$) and interruption times ($t_i$) on the growth of 2D InSe.** (**a**) Representative AFM images of the samples grown with different $t_d$ of (**a**) 10, (**b**) 19, and (**c**) 45 s with the fixed $t_i$ of 150 sec. The arrows in (a) and (b) indicate the 3D InSe islands. The inset of (c) shows the Raman spectrum of the corresponding film, indicating the growth of $In_2Se_3$ instead of InSe. (**d-f**) AFM characterizations of the InSe morphologies varied depending on the $t_i$ of (**d**) 120, (**e**) 150, (**f**) 180 s, while the $t_d$ was applied for 30 s. The pulsed injection was repeated 55 times for the study in (a-f). (**g**) Schematics depicting the lateral growth of 2D InSe depending on the $t_d$ and $t_i$. See Supplementary Note 1 for more descriptions of the growth aspects.

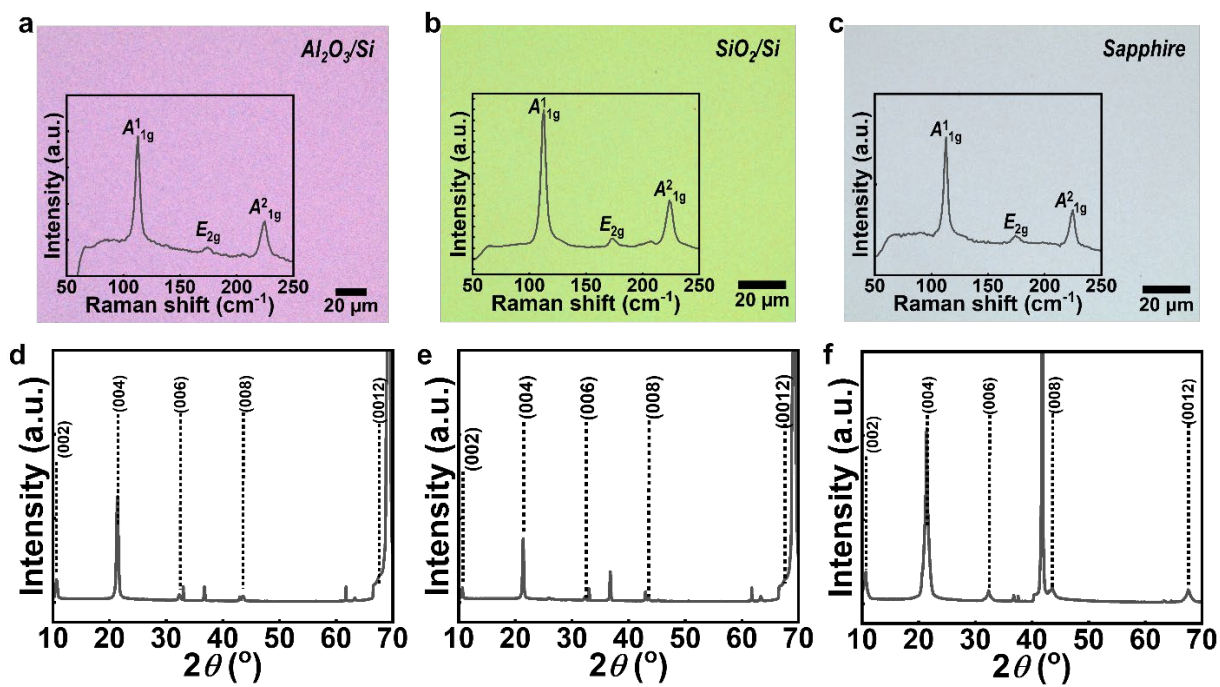

**Figure S5. Synthesis of (002*l*) textured InSe on the arbitrary substrates.** (**a-c**) OM images of the InSe film grown on (**a**) AlO$_x$/Si, (**b**) SiO$_2$/Si, and (**c**) *c*-plane sapphire. The insets in each image displayed their representative Raman spectrum of InSe. (**d-f**) XRD patterns of the InSe prepared by MOCVD on the corresponding substrates; (**d**) AlO$_x$/Si, (**e**) SiO$_2$/Si, and (**f**) *c*-plane sapphire. Each (002*l*) plane of InSe is indicated as dashed lines. All the peaks other than InSe originated from its substrates.

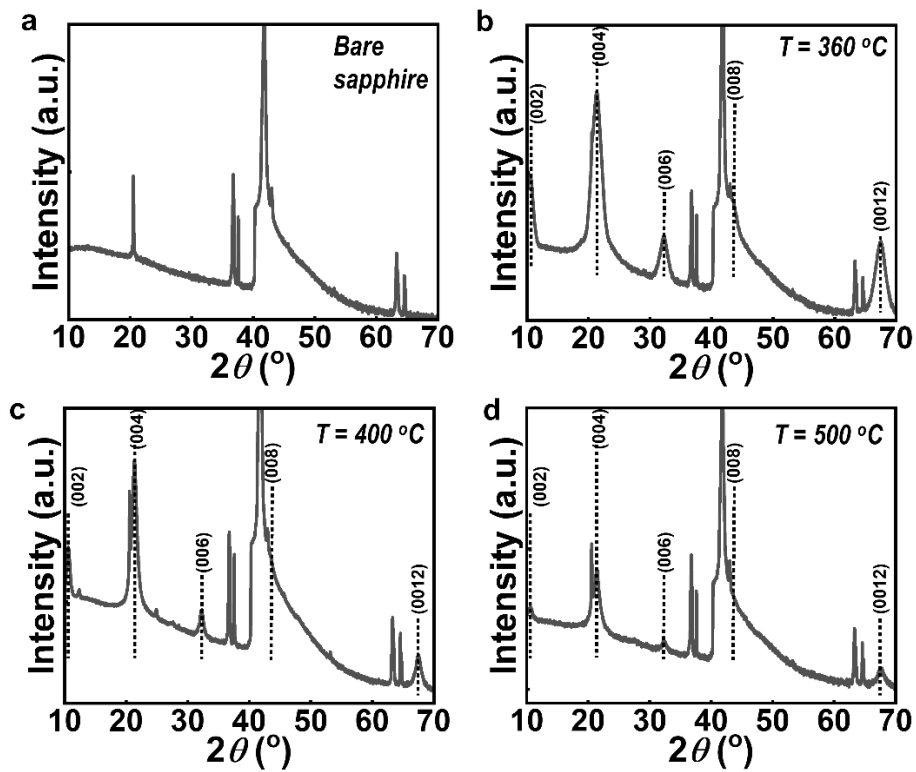

**Figure S6. XRD characterizations of InSe film on sapphire grown at different *T* plotted in a log scale.** (**a**) XRD pattern of bare *c*-plane sapphire. (**b-d**) XRD patterns of InSe grown at *T* of (**b**) 360, (**c**) 400, and (**c**) 500 °C, showing the (002*l*) peaks at the dashed lines.

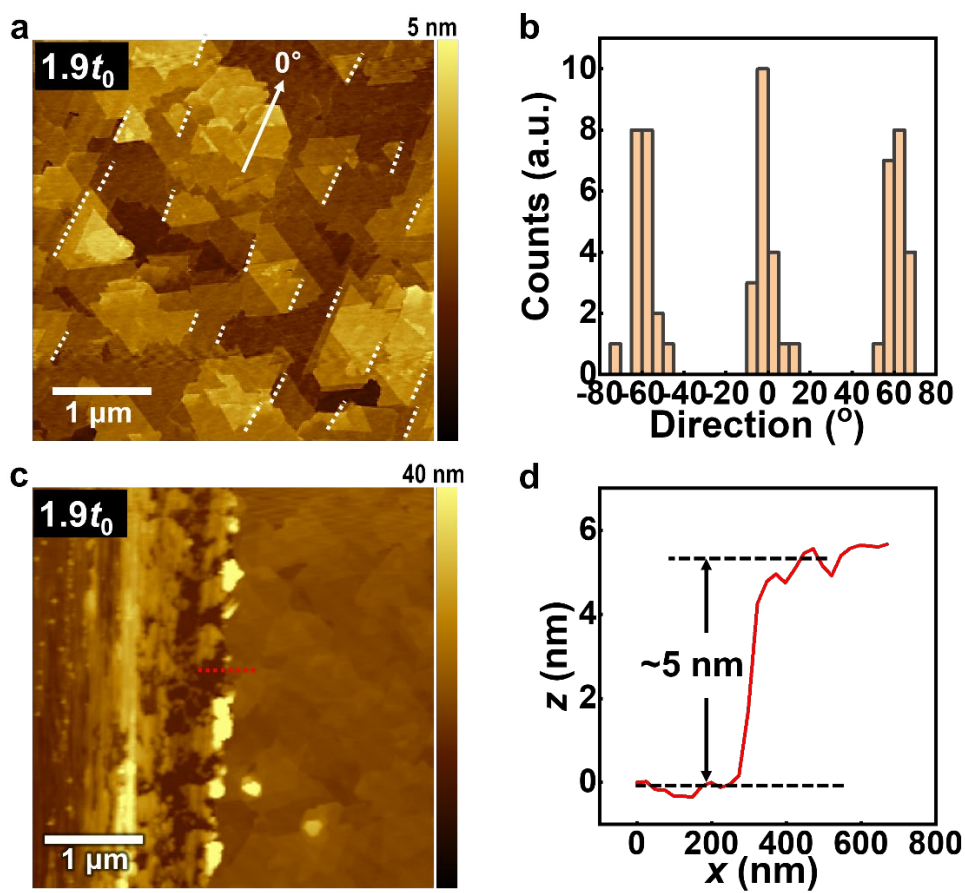

**Figure S7. AFM characterizations of InSe film grown for 1.9$t_0$.** (**a**) Representative topography AFM image of the InSe grown, showing the directionally aligned InSe domains. (**b**) Corresponding histogram of the orientation distribution of three edges of triangular InSe domain. The randomly selected 20 individual domains were counted; for example, one of the edges of InSe grains aligned to the direction of 0° is displayed as dashed lines in (a). (**c, d**) AFM analysis of the thickness of InSe film grown for 1.9$t_0$, (**c**) AFM image, and (**d**) the intensity profile along the dashed line in (b).

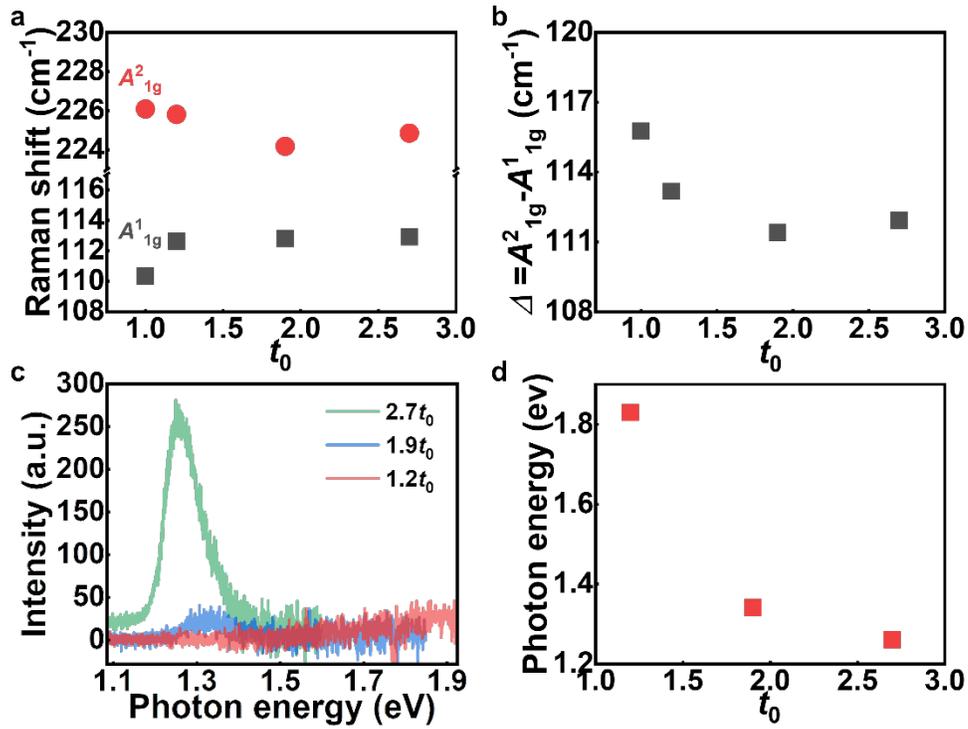

**Figure S8. Raman and PL analysis of the InSe grown for different *t*.** (**a**) Extracted peak positions for the Raman $A^2_{1g}$ and $A^1_{1g}$ vibrational modes of the InSe films. (**b**) Difference of the locations of the two $A_{1g}$ peaks as a function of *t*; e.g., $0.9t_0$, $1.2t_0$, $1.9t_0$, and $2.7t_0$. (**c**) Representative PL spectra of the InSe films with different *t*. The signals from sapphire substrates were subtracted from each plot. (**d**) Calculated photon energy emitted from each InSe film depending on the growth *t*. The extraction was conducted by Gauss fitting.

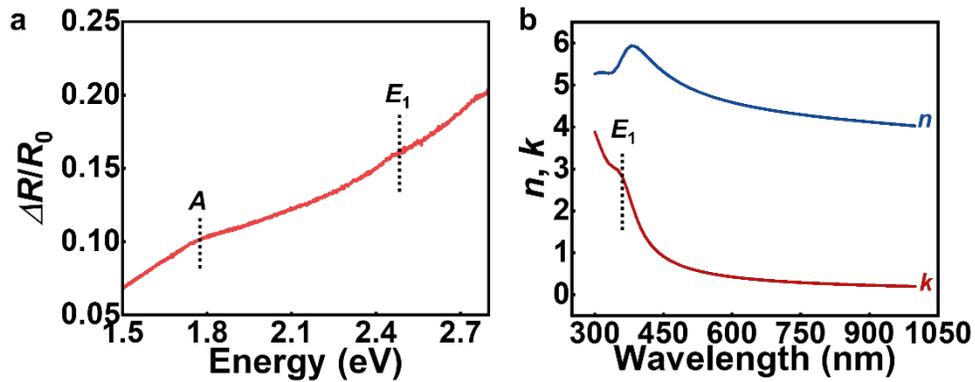

**Figure S9.** **Optical characterizations of monolayer InSe film.** (**a**) Differential reflectance ($\Delta R/R_0$) spectra of the monolayer InSe/sapphire film. (**b**) Real and imaginary parts of the refractive index; $\eta = n + ik$, where $n$ is directly related to propagation and power localization and $k$ is related to absorptivity/extinction. The bulk transition at ~3.1 eV was labeled as $E_1$, and its position was comparable with literature[2]. Spectroscopic ellipsometry (SE) was performed on as-grown films using a Woollam RC2 ellipsometer at incidence angles of 55°, 65°, and 75°. SE of bare sapphire surfaces was later conducted with a Woollam VASE ellipsometer at a 65° incidence angle. The complex permittivity was extracted from the raw Psi-Delta data by fitting the data with Lorentzian oscillators using the CompleteEASE program.

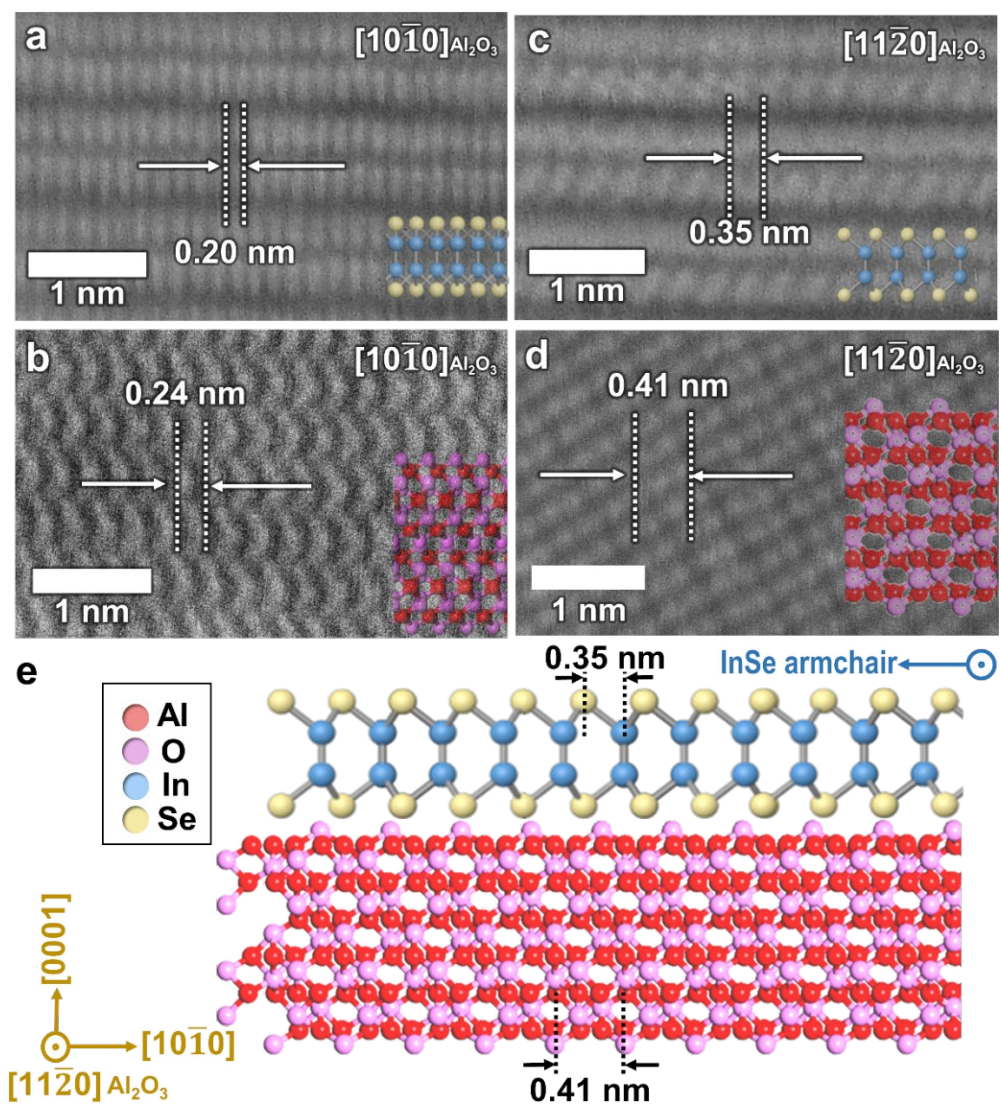

**Figure S10. Structural characterization of the cross-section of few-layer InSe grown on *c*-plane sapphire.** (**a-d**) STEM images of the InSe/sapphire showing the InSe planes sectioned along the (**a-b**) zigzag and (**c-d**) armchair directions. The STEM images in (a, c) and (b, d) were captured after focusing on InSe and $Al_2O_3$ surface, respectively, in Figure 3c, d. The InSe captured along the armchair direction demonstrated its characteristic atomic structure with broken mirror-plane symmetry (Figure S10c). (**e**) Schematic of InSe single layer and its atomic arrangements while demonstrating the epitaxial relation of $(0001)_{InSe}//(0001)_{Al_2O_3}$ and $(11\bar{2}0)_{InSe} //(10\bar{1}0)_{Al_2O_3}$.

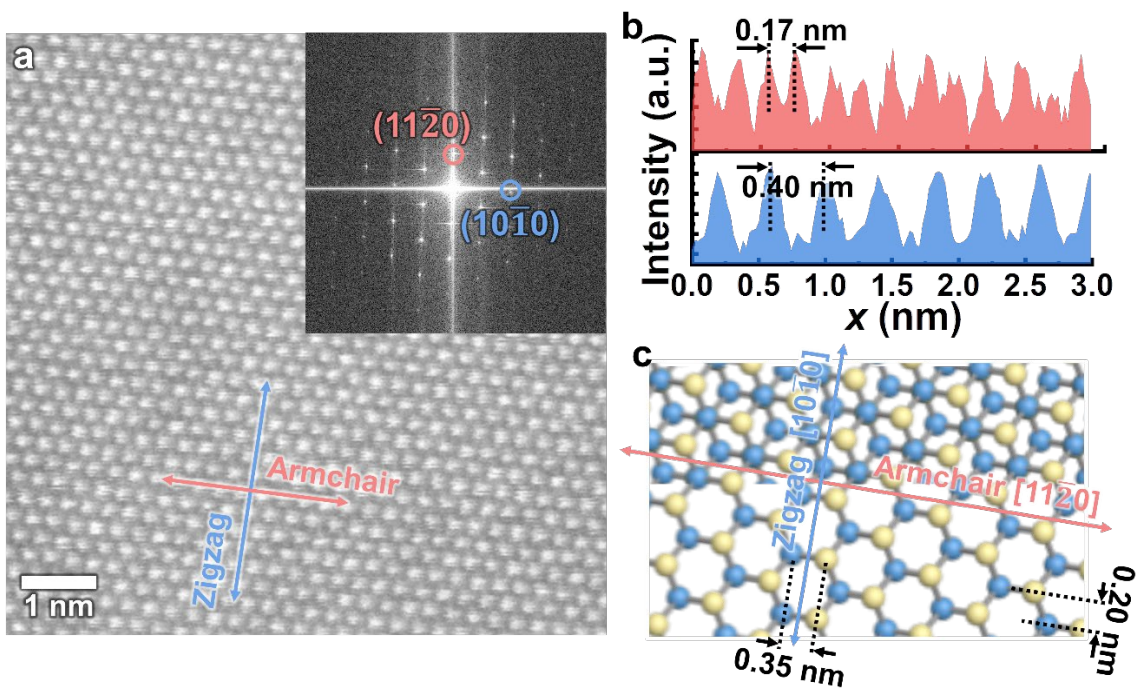

**Figure S11. Top-view STEM analysis of InSe.** (**a**) Atomic-resolution STEM image of few-layer InSe. The inset on the right corner is the corresponding diffractogram, indicating the planes aligned with a three-fold symmetry. (**b**) Intensity line profiles taken along the armchair and zigzag directions in (a) (**c**) Atomic schematic of the (bottom) InSe monolayer and (top) few layers. The same intensity of line profiles along the zigzag and armchair directions (Figure S11b) implicated that the InSe layers consisted of an AB stacking, not an AA-stacking[3], which suggested $\varepsilon$ or $\gamma$ phase of InSe (Figure S1c,d, and Figure S11c).

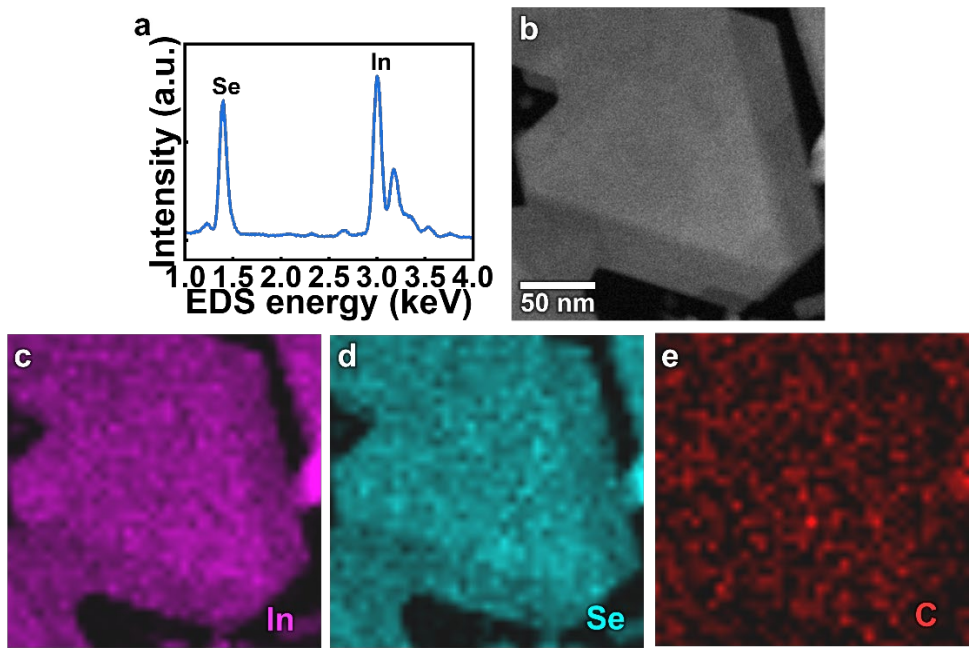

**Figure S12. TEM-EDS analysis of InSe crystal. (a)** Representative energy-dispersive X-ray spectroscopy (EDS) spectrum, indicating stoichiometric InSe with the estimated Se/In atomic ratio of ~1.00. (**b**) HAADF-TEM image of the InSe and its corresponding elemental mapping images for (**c**) In *L*, (**d**) Se *L*, and (**e**) C *K* signals. The uniform distributions of In and Se atoms in the crystal were indicated, whereas the C impurities from the InSe were not significant compared to the ambient.

The TEM-EDS investigations suggested the growth of high-quality InSe, as the at. % of Se/In of InSe (~1.00) estimated from the TEM-EDS spectrum was identical to the ideal value. The In and Se atoms were uniformly distributed along the few-layer InSe crystal with insignificant carbon signals in TEM-EDS mapping images. Similarly, the Raman spectra did not detect the features related to amorphous carbon (a-C) (Figure S13). The absence of a-C could be attributed to the repeated interruption of the precursor while flowing the $H_2$, where the process removed the unreacted carbon molecules[4,5].

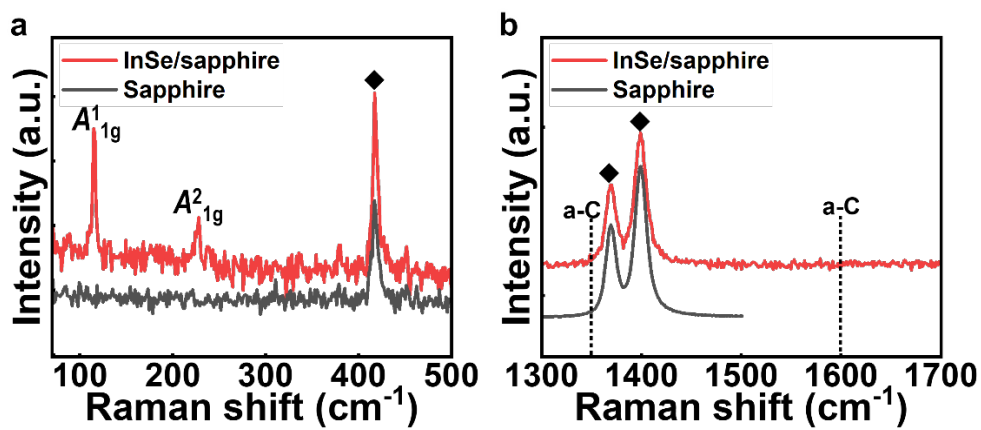

**Figure S13. Absence of amorphous carbon (a-C) in the produced InSe.** Raman spectra of InSe/sapphire (red) and bare sapphire (black) showing the (**a**) low- and (**b**) high-frequency regimes. Raman peaks for sapphire (diamonds) and a-C (dashed lines) are displayed for comparison.

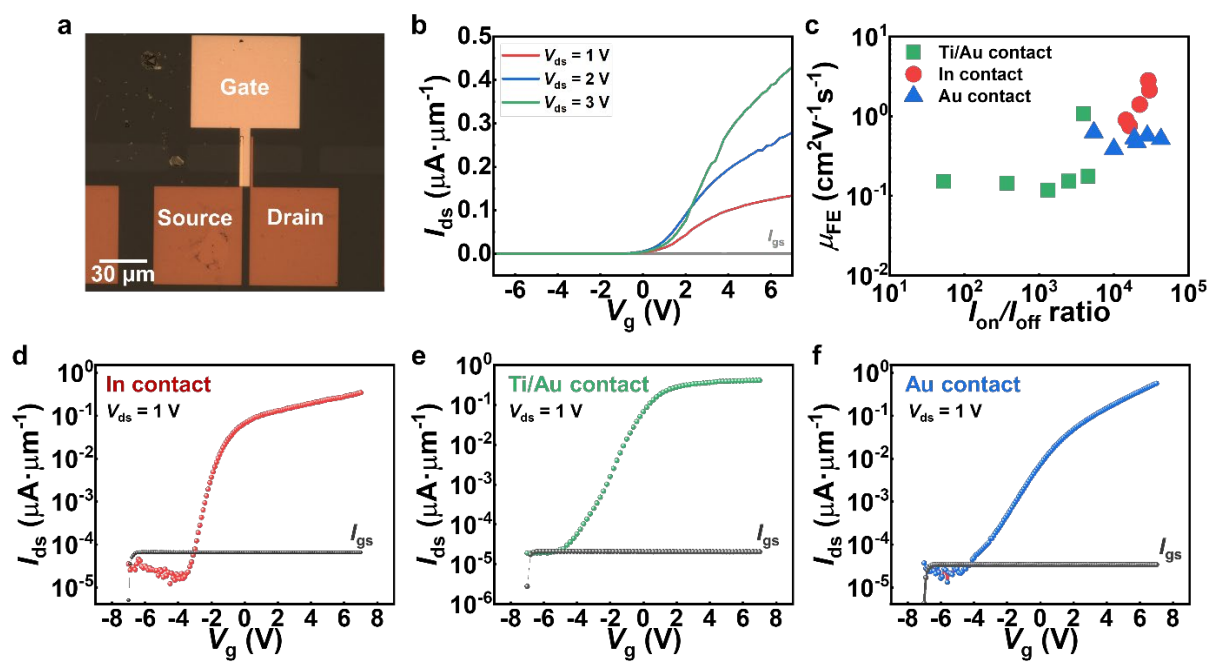

**Figure S14. Electrical properties of FETs based on few-layer InSe (~5 nm).** (**a**) OM image of a FET with top-gate dielectric and electrodes. (**b**) Representative transport curve of the InSe FETs under the different $V_{ds}$ of 1-3 V. (**c**) Comparisons of $\mu_{FE}$ and $I_{on}/I_{off}$ ratio of the InSe FETs with different contact electrodes. (**d-f**) Representative transfer curves of the device contacted with (**d**) In/Au, (**e**) Ti/Au and (**f**) Au electrodes measured at $V_{ds}$ of 1 V.

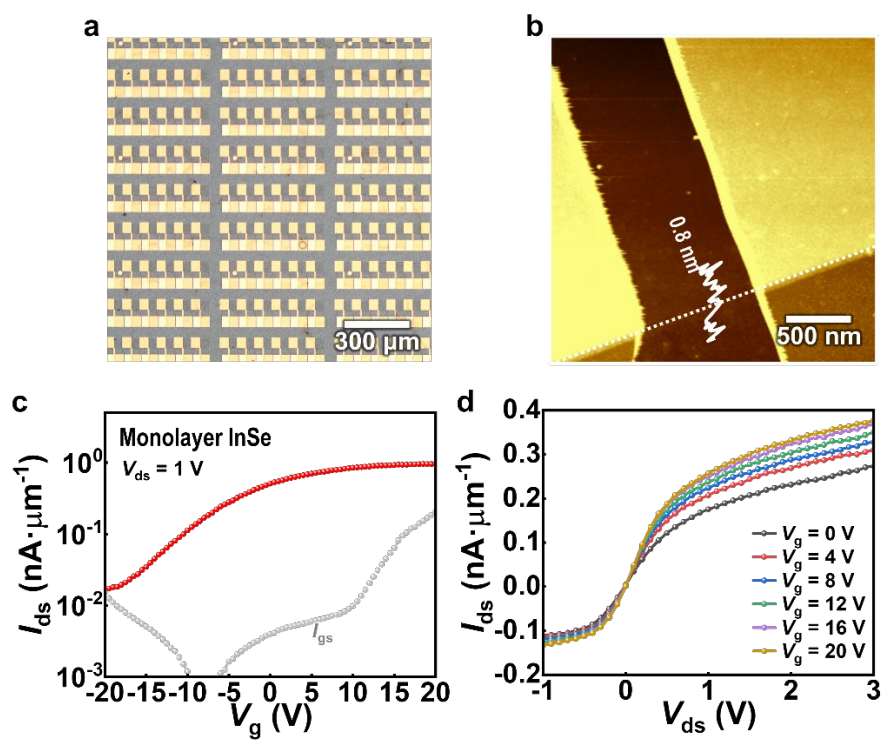

**Figure S15. Electrical characterization of monolayer InSe FETs.** (**a**) OM image of the fabricated FETs. (**b**) AFM image of the device showing the thickness of the monolayer InSe channel (scale bar: 500 nm). The image was captured before the deposition of gate dielectric and gate electrode. (**c**) Representative transfer curve of FETs based on monolayer InSe, when the drain voltage ($V_{ds}$) of 1 V was applied. (**d**) Output characteristic of monolayer InSe FET.

The AFM image of the defined InSe channel with ~0.8 nm thickness indicates the successful preparation of the devices with the monolayer (Figure S15b). Figure S12c demonstrates the drain-source currents ($I_{ds}$) of monolayer InSe device depending on the gate voltages ($V_g$) at the fixed drain voltage ($V_{ds}$) of 1 V. The two-terminal $\mu_{FE}$ of the monolayer FETs approached ~0.005-0.010 cm$^2$V$^{-1}$s$^{-1}$ while displaying the $I_{on}/I_{off}$ ratio of ~10$^2$, which values are comparable or even higher than the previous reports for the FETs based on exfoliated[6] and CVD-grown[7] InSe monolayer flakes. Since the transport of the monolayer can be highly degraded by the substrate trap density and oxidation during the device fabrication, its performance is poor compared to the thicker InSe[6,7]. Accordingly, the output characteristics ($I_{ds}$-$V_{ds}$) of monolayer InSe exhibited modest $I_{ds}$ differences depending on $V_g$ even after the current saturation (Figure S15d).

**Table S1. Comparisons of preparation methods for few-layer InSe and their electrical properties characterized by two-terminal FETs with dielectric layers.** The "1L" in Thickness indicates a monolayer of InSe.

| | InSe preparation methods | Thickness | Lateral size of the prepared InSe | $\mu_{FE}$ ($cm^2V^{-1}s^{-1}$) | $I_{on}/I_{off}$ ratio | Contact metals | Dielectric layer | Reference |
|---|---|---|---|---|---|---|---|---|
| Direct synthesis | MOCVD | 1L | 5.08 cm (2 inches) | 0.01 | $10^2$ | In/Au | 30 nm $AlO_x$ | This study |
| | | 5 nm | | 2.81 (average: 1.0 ± 0.8) | $10^4$-$10^5$ | | 10 nm $Sb_2O_3$ + 10 nm $HfO_x$ | |
| | CVD | 1L | few μm | $3*10^{-5}$ | $10^{0.3}$ | Au | $SiO_2$ | 7 |
| | | 3-100 nm | | N/A | | | | 8 |
| | PLD and post-annealing | 1 nm | 1 cm | 10 | $3*10^2$ | Au | 300 nm $SiO_2$ | 9 |
| | | 15 nm | | 0.2 | ~$10^4$ | Au | 3 nm $AlO_x$ + 20 nm $HfO_x$ | 10 |
| | CVT | Few layers | few μm | 10.8 | $10^{0.84}$ | Cr/Au | 300 nm $SiO_2$ | 11 |
| | PVD | Few layers | | N/A | ~$10^3$ | Cr/Au | 285 nm $SiO_2$ | 12 |
| Exfoliation | Mechanical exfoliation | 1L | few μm | 0.02 | ~$10^2$ | Cr/Au | hBN/$SiO_2$ | 6 |
| | | 6L | | N/A | ~$10^5$ | | | |
| | | 2L | | 0.4 | ~$10^4$ | Au (vdW contact) | hBN/$SiO_2$ | 13 |
| | | 5L | | 3 | ~$10^5$ | | | |
| | | 9L | | 20 | ~$10^6$ | | | |
| | | 12 nm | | 0.1 | ~$10^4$ | Cr/Au | 300 nm $SiO_2$ | 14 |
| | Chemical exfoliation | 30 nm | | 2 | ~$10^4$ | Ti/Au | $SiO_2$ | 15 |

**Supplementary Note 1. The roles of DMSe interruption for 2D InSe growth**

To investigate the effects of repeated interruption/delivery of DMSe on growth aspects, the MOCVD is conducted with different Se/In ratios under either constant or various DMSe flows (Figures S2 and S4, respectively).

The structural properties of as-grown InSe and $In_2Se_3$ crystals are explored by using Raman spectroscopy and AFM depending on the Se/In ratios under the constant delivery of DMSe (Figure S2). The total flow is almost constant (~50 sccm) for this investigation. The Se/In ratio from ~3 to 66 promotes the growth of $\beta$-$In_2Se_3$, whereas the Se/In ratio of ~2 yields the production of InSe, as confirmed by the Raman spectra (Figure S2a). Considering the binary In-Se system (Figure S1), the growth of InSe under the Se-deficient ambient is expected because it is thermodynamically stable. Nonetheless, the resultant InSe structure has a spherical-shaped island instead of an atomically flat, 2D morphology (Figure S2b). The high In flux triggers the more chances of In absorption on the substrate, resulting in the coalescence of liquid In droplets at the growth temperature (~500 °C) while reacting with Se to produce InSe.[16,17] The formation of In or Ga droplets is also a common phenomenon during MOCVD[18] and MBE[16] of III-nitride compounds caused by the smaller V/III ratio.

In contrast, the higher Se/In ratio up to ~11 causes the layer-by-layer growth of $In_2Se_3$ flakes, as the triangular domains and their one monolayer thickness of ~1.0 nm are observed in the AFM image (Figure S2c). Under the high chalcogen-to-metal ratio, the layer-by-layer growth can be triggered during MOCVD while increasing the domain size, as previously studied for 2D $In_2Se_3$[19] or 2D group-VI TMDs[20], because sufficient Se atoms on the surface prevents the formation of elemental metal clusters and 3D growth. This indicates that the production of 2D InSe layers with larger domain sizes is exceedingly challenging just by changing the Se/In ratio.

On the other hand, the pulsed mode for DMSe (i.e., $t_d/(t_i+t_d) < 1$) allows the layer-by-layer growth of 2D InSe even under the maximum Se/In ratio up to ~11 (Figure 1e). In addition, the AFM images in Figure 2d-f show the successful growth of the triangular- or hexagonal-like shaped InSe domains with a thickness of ~0.8 nm, indicating the as-grown structures are the 2D monolayer. The $c$-plane-orientated InSe thin film is confirmed by its X-ray diffraction pattern, which also supports layer-by-layer growth (Figures S5 and S6). To study how the repeated DMSe delivery and interruption affect the growth of 2D InSe, the times for each periodic flow; i.e., $t_d$ and $t_i$, are changed and the as-grown structures' morphologies are investigated by AFM, as shown in Figure S4. The increased $t_d$ from 10 to 45 s for the fixed $t_i$ of 120 s reduces the densities of 3D InSe; however, when the $t_d$ exceeds 45 s, the synthesized crystal is $In_2Se_3$ instead of InSe, where the phase transition can be resulted from higher Se flux by longer $t_d$ (Figure S4a-c). On the other hand, when the $t_d$ is not changed (30 s), the increase in $t_i$ from 120 s to 150 s leads to the larger size domains with sharp edges, whereas much longer $t_i$ of ~180 s induces the degradation of structures (Figure S4d-f).

Considering the growth results in Figure S4a-f, the possible mechanism of the growth of 2D InSe layers is suggested as follows: When the delivery of DMSe flow is interrupted (introduction of $t_i$), the relatively deficient Se atmosphere allows the nucleation of InSe instead of $In_2Se_3$. During this period, the ripening of the nucleus and lateral growth also happens due to the increased diffusion of $In_xSe_y$ clusters absorbed on the substrate[21,22]. On the other hand, the introduction of Se flux (an increase of $t_d$) prevents the formation of In-rich clusters or In droplets to obtain pure 2D InSe layers instead of InSe islands. Consequently, the repeatedly varied atmospheres by pursuing of DMSe source permit not only nucleation of InSe but also ripening and lateral growth of the domain. The exceedingly long $t_d$ and $t_i$ can arouse either phase transition of InSe to $In_2Se_3$ (by high Se/In ratio) or degradation of the crystals (by small high Se/In ratio); thereby, the optimizations of $t_i$, $t_d$, and Se/In ratio are highly encouraged to attain 2D InSe atomic layers with high crystallinity.